# Thallium Strontium Iodide: A High Efficiency Scintillator for Gamma-ray Detection

Rastgo Hawrami[1,2], Elsa Ariesanti[1], Vlad Buliga[1], Arnold Burger[1]

*[1]Fisk University, Nashville, TN 37208*
*[2]Xtallized Intelligence, Inc., Nashville, TN 37211*

### *Abstract*

Europium-doped $TlSr_2I_5$ (TSI), a new high light yield scintillator for gamma-ray detection, has been grown by the vertical Bridgman method. A 12mm and 16mm diameter boules of TSI is grown in a two-zone vertical furnaces. Samples extracted from the grown boule have been characterized for their scintillation properties. Energy resolution of < 3% (FWHM) at 662 keV and a gamma-ray light yield of approximately 54,000 Ph/MeV have been obtained. Decay times of 395 ns (89%) and 2.0 µs (11%) have been measured.

### *Keywords*

Tl-based scintillators. Dense high Z scintillator. Advanced scintillator. High efficiency scintillator. Excellent energy resolution. High light yield scintillator.

### *Highlights:*

- Europium-doped $TlSr_2I_5$ (TSI) has been grown by the vertical Bridgman method.
- Energy resolution of <3% (FWHM) at 662 keV has been measured.
- Gamma-ray light yield of approximately 54,000 ph/MeV has been obtained.
- Adding Tl to or substituting Tl for other (metal) alkali component in the compounds improves scintillation performance.

## 1. INTRODUCTION

Demands for high light yield, high density, and fast scintillators necessitate a continuous search for new materials. Traditional scintillators such as Tl-doped sodium iodide (NaI) and cesium iodide (CsI) have been very reliable standards, supported by decades of research and proven performance. However, various new applications require bright materials that also have high densities and fast decay times. For nearly two decades emerging new scintillators such as $LaX_3$ [1-3], $CeX_3$ [4, 5], and $Cs_2AX_5$ [6], where A = La or Ce, and X = Cl, Br, or I (halides), have demonstrated the potentials of these metal halides as next-generation scintillation detectors. Rediscovered of europium-doped $SrI_2$ with a light yield as high as 110,000 ph/MeV and moderate density of 4.55 $g/cm^3$ has also shown the potential of alkaline metal halide scintillators [7, 8].

Recently Tl-based scintillators have attracted good attention from worldwide scintillator researchers and have given rise to exploration of new scintillation materials in a new direction. These compounds have been investigated and very promising initial results have been published. Tl-based metal halides such as intrinsic scintillators $TlMgCl_3$ (TMC) and $TlCaI_3$ (TCI), as well as Ce-doped $Tl_2LaCl_5$ (TLC) have also been investigated and yielding good results [9,10]. Table 1 shows these several recently reported Tl-based scintillation crystals and their respective lanthanum or alkaline earth halides [1-3,7-12]. Note that undoped $MgCl_2$ has not been found as a practical scintillator and is listed in Table 1 only to demonstrate the effect of Tl-ion addition to density and $Z_{eff}$. These new compounds are of high densities (< 5 $g/cm^3$), bright (light yields between 31,000 and 76,000 ph/MeV for 662 keV photons), fast decay times (36 ns (89%) for TLC; 46 ns (9%) for TMC; 62 ns (13%) for TCI), and moderate melting points (between 500 and 700ºC). As seen further in the published results, Tl-based scintillators such as the ones previously mentioned have promising properties desirable for high energy physics as well as homeland security applications.

Table 1 shows that adding Tl to or substituting Tl for other (metal) alkali component in the compounds increases the density and $Z_{eff}$ and improves scintillation performance. For example, in the case of SrI$_2$ [11,12] and TlSr$_2$I$_5$ [12] material density increases from 4.55 to 5.30 g/cm$^3$, $Z_{eff}$ increases from 49 to 61, and the primary decay time of TlSr$_2$I$_5$ is shorter than SrI$_2$ by about 0.5. TlSr$_2$I$_5$ with 3% EuI$_2$ doping was recently investigated and reported in [7] and found to have good radiometric properties such as 4% energy resolution at 662 keV, as well as a high light yield of 70,000 ph/MeV [7]. During the same time of the study in [12], we independently investigated and grew intrinsic TlSrI$_3$ and 5% Eu-doped TlSr$_2$I$_5$ (TSI).

Table 1. Recently reported Tl-based scintillation crystals and their respective lanthanum or alkaline earth halides.

| Scintillator | Energy Resolution at 662 keV (%) | Light Yield (Ph/MeV) | Decay time (ns) | Peak Emission (nm) | $Z_{eff}$ | Density (g/cm$^3$) | Ref. |
|---|---|---|---|---|---|---|---|
| LaCl$_3$:Ce | 4.00% | 46,000 | 25 | 350 | 59.5 | 3.79 | [1], [2] |
| Tl$_2$LaCl$_5$:Ce | 3.40% | 76,000 | 36 | 383 | 70.2 | 5.2 | [10] |
| LaBr$_3$:Ce | 3.00% | 63,000 | 20 | 380 | 48.3 | 5.29 | [3] |
| Tl$_2$LaBr$_5$:Ce | 6.30% | 43,000 | 25 | 435-415 | 67 | 5.9 | [11] |
| MgCl$_2$ | N/A | N/A | N/A | N/A | 16.1 | 2.32 | N/A |
| TlMgCl$_3$ | 3.70% | 36,000 | 46 ,166, 449 | 409 | 69.7 | 5.3 | [9] |
| SrI$_2$:Eu | 3% | 85000 | 1200 | 435 | 50 | 4.6 | [7], [8] |
| TlSr$_2$I$_5$:Eu | 4% | 70,000 | 525 | 463 | 61 | 5.3 | [12] |

For more than a decade researchers currently based at Fisk University have investigated and developed SrI$_2$:Eu as an excellent candidate for scintillation detector to replace LaBr$_3$:Ce [3]. SrI$_2$:Eu has excellent scintillation properties, with current best energy resolution of 2.7% at 662 keV, a light yield as high as 110,000 Ph/Mev, and good non-proportionality [7,8]. As shown in Table 1 adding Tl to SrI$_2$ is an attractive way to increase stopping power while preserving SrI$_2$ excellent scintillation properties. Using total mass attenuation coefficient data calculated by NIST XCOM online program [13] and using published mass density values for the compounds (see Table 1) for NaI, SrI$_2$ and TlSr$_2$I$_5$, the linear attenuation coefficients for these materials can be calculated and are shown in Fig. 1(a). At low photon energy, the photoelectric effect dominates, and the attenuation coefficient shows the effects of absorption at the electron shell edges. Over a wide range of photon energy, TlSr$_2$I$_5$ has a higher attenuation coefficient than SrI$_2$ and NaI. To show the effect of a higher attenuation coefficient, fraction of parallel beam of photons attenuated (i.e., undergone interactions) in a 1 cm thick infinite slab of material was calculated. Fig. 1(b) clearly shows that the higher the attenuation coefficient, the more interactions can occur in a material. In other words, for the same material thickness and for any gamma-ray energy, more radiation is attenuated by TlSr$_2$I$_5$ than by either NaI or SrI$_2$. Tl-based compounds demonstrate that they have higher absorption or attenuation coefficients than the original lanthanide and alkali metal halides, thus more interactions can occur in Tl-based scintillators.

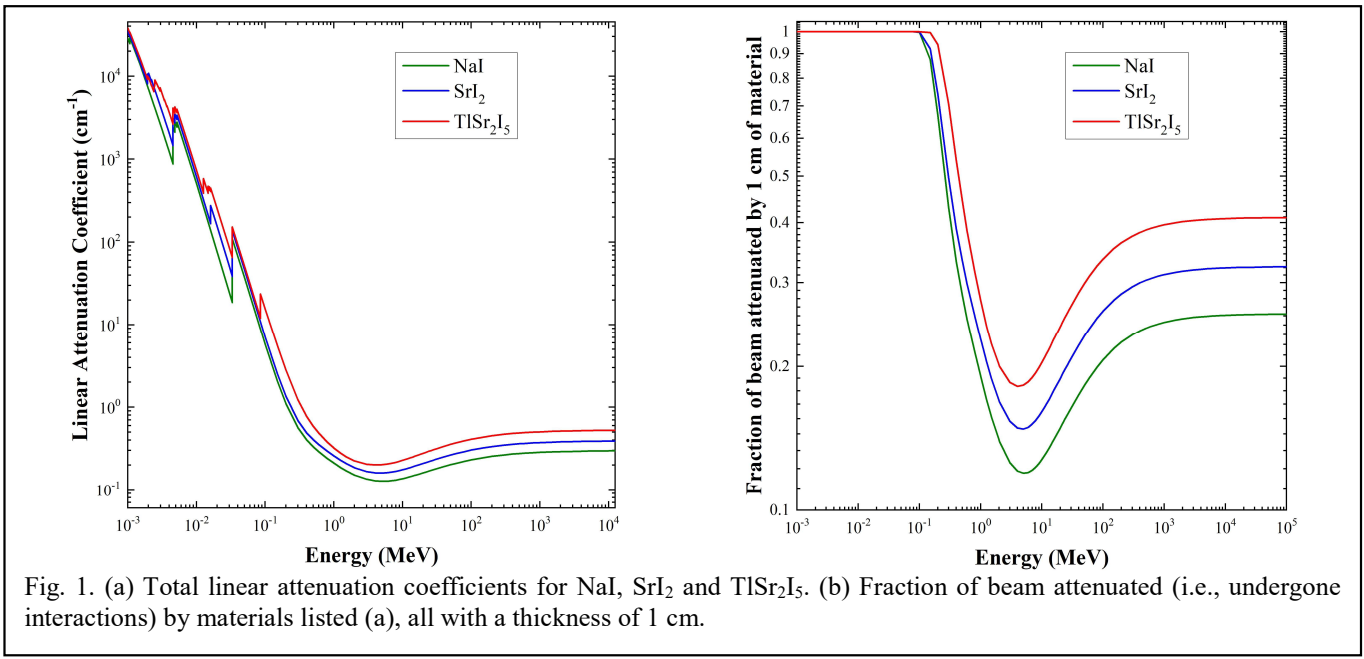

Fig. 1. (a) Total linear attenuation coefficients for NaI, SrI$_2$ and TlSr$_2$I$_5$. (b) Fraction of beam attenuated (i.e., undergone interactions) by materials listed (a), all with a thickness of 1 cm.

TSI scintillator has monoclinic crystal structure with lattice parameters a = 9.97 Å, b = 8.94 Å, and c = 14.25 Å [14]. The space group of this scintillator is P21/C and density 5.30 g/ cm$^3$. Effective Z-number was found to be 61 [12]. Luminescence properties of the grown crystals are measured under X-ray excitation at room temperature. Pure and doped crystals contained broad emission bands between 445-670 nm peaking at 528 nm and 430-600 nm peaking at 463 nm, respectively [12]. In this publication we are reporting the vertical Bridgman growth and characterization of 5% Eu-doped TlSr$_2$I$_5$ (TSI). More physical and thermal characteristics will be explored and published in next a following paper including currently investigated new Tl-based compounds.

## 2. Experimental Methods

Stoichiometric amounts of TlI, SrI$_2$, and EuI$_2$, all in powder form, were placed in a clean ∅12mm and 16 mm inner diameter quartz ampoules, for which each was subsequently sealed under high vacuum and placed in a two-zone vertical Bridgman furnace. Furnace zone temperatures were set to facilitate melting at around 500ºC. Crystal growth commenced at a rate of 10 mm/day and post-crystallization cooling at a rate of 100ºC/day. Fig. 2 shows the as-grown boule of TSI. Samples retrieved from the boule were lapped and polished with sandpapers. Mineral oil was used for lubrication and sample protection from moisture because TSI, like SrI$_2$, is hygroscopic. The polished samples were tested for their radiometric and scintillation properties. For each sample measurement, the sample was placed in mineral oil in a quartz cup wrapped with Teflon tape as a reflector. A piece of Gore® flexible Teflon sheet was used as the back reflector. Using optical grease, the oil cup was coupled to a R6231-100 Hamamatsu super bi-alkali photomultiplier tube (PMT), coupled to standard nuclear instrument module (NIM) equipment (Hamamatsu C9525-02 high voltage power supply, Canberra 2005 preamplifier, Canberra 2020 spectroscopy amplifier, and Amptek MCA800D multichannel analyzer). $^{137}$Cs spectrum was collected to obtain energy resolution and was compared to $^{137}$Cs spectrum collected with a calibrated ∅1"×1" NaI:Tl to obtain light yield for TSI. Then spectra from check sources with x-/γ-ray energies between 14 keV and 1332 keV ($^{241}$Am, $^{57}$Co, $^{22}$Na, and $^{137}$Cs) were collected to obtain energy resolution and light yield at specific photon energies. Relative light yield data were obtained by normalizing the light yield at the specific energy to that of $^{137}$Cs. The non-proportionality curve was obtained by plotting the relative light yield as a function of photon energy. To obtain luminescence decay time information waveforms from the PMT anode were recorded with CAEN DT5720C digitizer and analyzed offline. For spectra measurement, the power supply was set at -700V, while for scintillation decay measurement, the power supply was set at -1000V. The emission spectrum was collected using a modular system from Ocean Optics, with a DH-2000-BAL light source with an inline linear variable filter (300-750 nm range) and a FLAME-S spectrometer. Data were collected using the OceanView® software and analyzed off-line.

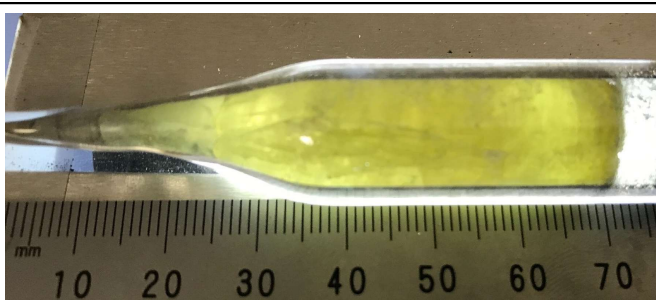

Fig. 2. A photograph of a boule of TlSr$_2$I$_5$:Eu (TSI) grown with the vertical Bridgman technique.

## 3. RESULTS AND ANALYSIS

### 3.1. Emission Spectrum

The emission spectrum collected by a TSI sample from the 16 mm boule (see inset picture) is plotted in Fig. 3. Emission peak at about 420 nm was measured.

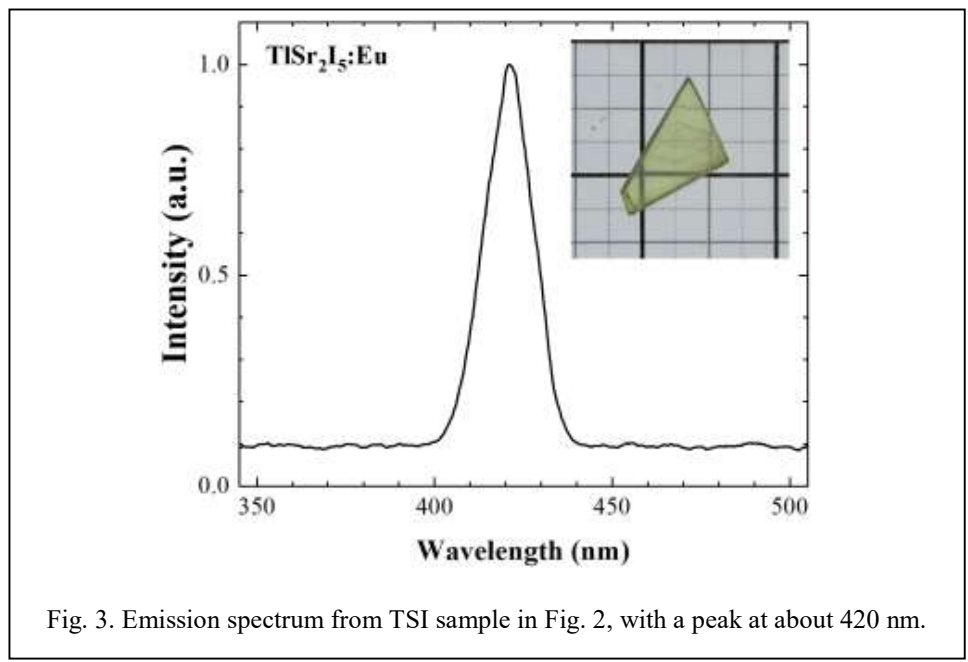

Fig. 3. Emission spectrum from TSI sample in Fig. 2, with a peak at about 420 nm.

### 3.2. Energy Resolution and Light Yield at 662 keV

Figs. 4(a) and (b) show [137]Cs spectra collected by two different TSI samples shown in the inset pictures ($10 \times 8.5 \times 2$ mm$^3$ in (a) and $14 \times 11 \times 9$ mm$^3$ in (b)) with 8 μs amplifier shaping time. Energy resolutions of 2.6% and 2.9% (FHWM) at 662 keV were calculated, respectively. Tl x-ray escape peaks is visible at a channel number (i.e., energy) below the full energy peak in each spectrum. The energy resolution measured in this study is better than previously reported, which was 4% at 662 keV for TlSr$_2$I$_5$:3%Eu [12]. Also shown in Fig. 4(a) is a comparison of [137]Cs spectra collected by the aforementioned TSI and a $\varnothing 1'' \times 1''$ NaI:Tl crystal, which was measured to have a light yield of 35,000 ph/MeV. Taking into account the emission peaks for TSI and NaI:Tl, respectively, the light yield of TSI was calculated to be approximately 54,000 ph/MeV. This light yield value is lower than previously reported, which was 70,000 ph/MeV. Dopant concentrations have been shown to affect light yield in many scintillating including Tl-based compounds [12]. The discrepancy in light yield values may be caused by the difference in Eu$^{2+}$ concentrations.

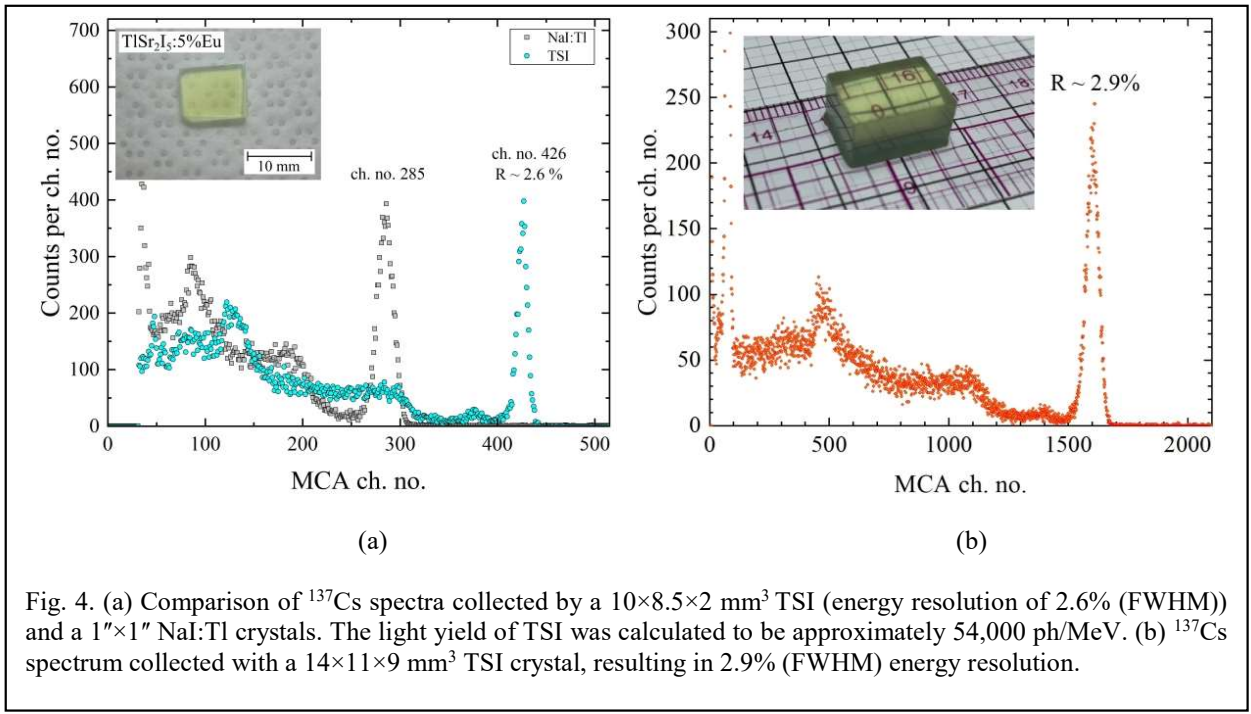

(a)                                                                      (b)

Fig. 4. (a) Comparison of $^{137}$Cs spectra collected by a 10×8.5×2 mm$^3$ TSI (energy resolution of 2.6% (FWHM)) and a 1″×1″ NaI:Tl crystals. The light yield of TSI was calculated to be approximately 54,000 ph/MeV. (b) $^{137}$Cs spectrum collected with a 14×11×9 mm$^3$ TSI crystal, resulting in 2.9% (FWHM) energy resolution.

### 3.3. Non-Proportionality

Fig. 5 shows the gamma-ray non-proportionality curves (i.e., relative light yield as a function of photon energy) for TSI for photon energies between 14 and 1275 keV. When compared to NaI:Tl and BGO, TSI has a fairly linear response to energies above 100 keV.

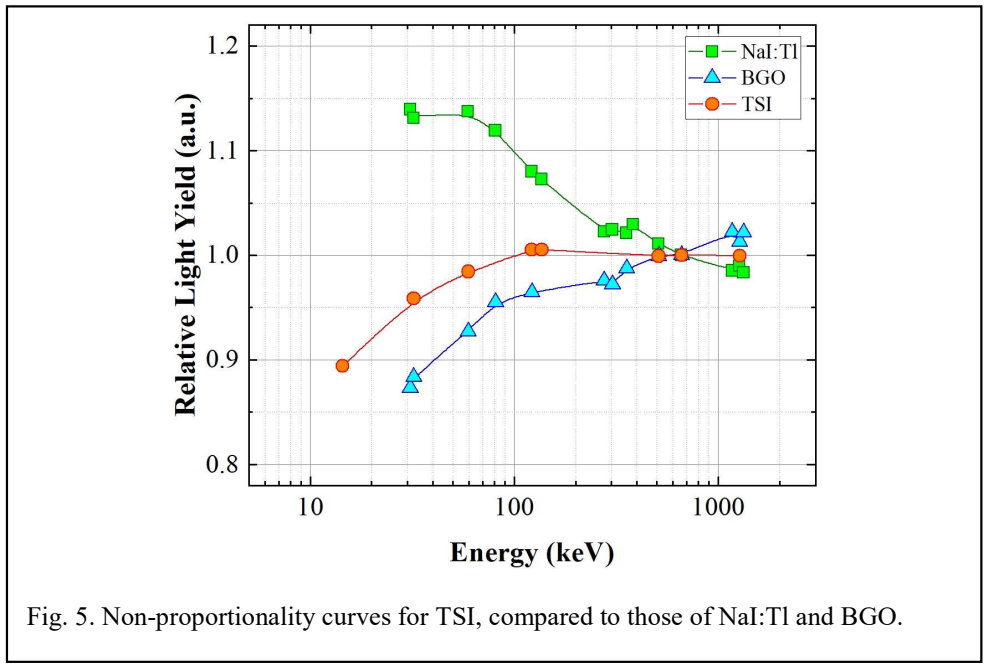

Fig. 5. Non-proportionality curves for TSI, compared to those of NaI:Tl and BGO.

### 3.4. Decay Time

Fig. 5 shows the averaged waveforms or signals collected at the PMT anode of the TSI sample in Fig. 2 irradiated with $^{137}$Cs. The exponential fit to the average curve yields a primary decay time of 395 ns (89%) and a longer, secondary decay time of 2.0 µs (11%). Dopant concentrations have been shown to affect decay times in many scintillating including Tl-based compounds, thus, further study of $Eu^{2+}$ levels in TSI may be needed to determine the optimal concentration.

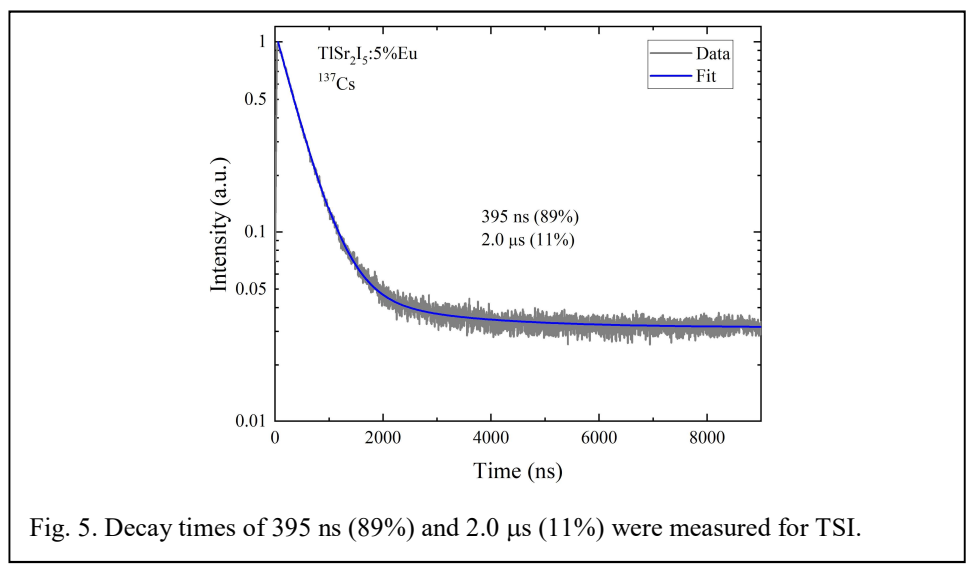

Fig. 5. Decay times of 395 ns (89%) and 2.0 µs (11%) were measured for TSI.

## 4. Conclusions

We are reporting a successful crystal growth of high scintillating performance of europium-doped $TlSr_2I_5$ (TSI), which is an attractive new bright scintillator for room-temperature gamma-ray detection. Preliminary results have shown that an excellent energy resolution below 3% at 662 keV is achievable. Moreover, TSI is faster than $SrI_2$:Eu. Because of the presence of Tl in the structure, its effective atomic number ($Z_{eff}$) and density are higher than $SrI_2$, which makes a higher stopping power for TSI. Furthermore, detection efficiency is directly proportional to density, which means denser materials have higher detection efficiency. The denser the material, the less volume is needed, resulting in lower production cost, thus making TSI an excellent candidate material for best radioisotope identification devices in homeland security applications and high energy physics applications.

## 5. Acknowledgment

This work was partially supported by US National Science Foundation under Grant #HRD-1547757.